\documentclass[twoside,twocolumn]{article}

\usepackage{tabulary,graphicx,times,caption,fancyhdr,amsfonts,amssymb,amsbsy,latexsym,amsmath}
\usepackage[utf8]{inputenc}
\usepackage{url,multirow,morefloats,floatflt,cancel,tfrupee,textcomp,colortbl,xcolor,pifont}
\usepackage[nointegrals]{wasysym}
\makeatletter

\usepackage{ifxetex}
\ifxetex\else
  \usepackage{dblfloatfix}
\fi

\@ifundefined{subparagraph}{
\def\subparagraph{\@startsection{paragraph}{5}{2\parindent}{0ex plus 0.1ex minus 0.1ex}%
{0ex}{\normalfont\small\itshape}}%
}{}

\def\URL#1#2{\@ifundefined{href}{#2}{\href{#1}{#2}}}

\def\UrlOrds{\do\*\do\-\do\~\do\'\do\"\do\-}%
\g@addto@macro{\UrlBreaks}{\UrlOrds}

\makeatother

\usepackage[paperheight=11in,paperwidth=8.3in,margin=2.5cm,headsep=.7cm,top=2.5cm]{geometry}
\usepackage[T1]{fontenc}

\renewenvironment{abstract}
	{\trivlist\item[]\leftskip0pt\par\vskip4pt\noindent
  	\mbox{\null}\\}
	{\par\noindent\endtrivlist}

 \date{} \emergencystretch 8pt

\makeatletter
\def\author#1{\gdef\@author{\hskip-\tabcolsep%
	\parbox{\textwidth}{\raggedright\bfseries#1\\[1pc]}}}
\def\address[#1]#2{\g@addto@macro\@author{\\\hskip-\tabcolsep\parbox{\textwidth}{\raggedright%
	\normalsize\normalfont\textsuperscript{#1}#2}}}
\let\addresslink\textsuperscript
\def\correspondence#1{\g@addto@macro\@author{\\\hskip-\tabcolsep\parbox{\textwidth}{\raggedright%
	\vspace*{10pt}\normalsize\normalfont~\\#1~\\[12pt]}}}
\def\email#1{\g@addto@macro\@author{\\\hskip-\tabcolsep\parbox{\textwidth}{\raggedright%
	\normalsize\normalfont Emails: #1}}}

\def\title#1{\gdef\@title{\vspace*{-30pt}%
	\raggedright\textbf{\@journaltitle}~\\%
  \raggedright\bfseries\ifx\@articleType\@empty\vspace*{20pt}\else%
  \vspace*{20pt}\@articleType\vspace*{20pt}\\\fi#1}}
\let\@journaltitle\@empty \def\journaltitle#1{\gdef\@journaltitle{{\normalfont\itshape#1}}}
\let\@articleType\@empty \def\articletype#1{\gdef\@articleType{{\normalfont\itshape#1}}}

\let\@runningHead\@empty \def\RunningHead#1{\gdef\@runningHead{{\normalfont #1}}}

\usepackage{fancyhdr}
\fancypagestyle{headings}{\fancyhf{}
  \fancyhead[R]{\itshape\@runningHead}
  \fancyfoot[C]{\thepage}}
\fancypagestyle{plain}{%
	\fancyhf{}\fancyhead[R]{}
  \fancyfoot[C]{\thepage}}
\makeatother

\usepackage[%
	numbers,sort&compress%
  ]{natbib}

\usepackage{float,xcolor}

\journaltitle{ }
\articletype{Research Article} 

\begin{document}

\title{Analytical model of low mass strange stars in 2+1 spacetime}

\author{Masum Murshid\addresslink{1},
  	Nilofar Rahman\addresslink{1}, 
  	Irina Radinschi\addresslink{2} and
  	Mehedi Kalam\addresslink{1}
    }
		
\address[1]{Department of Physics, Aliah University, IIA/27, New Town, Kolkata 700160, India}
\address[2]{Department of Physics, Gh. Asachi Technical University, Iasi 700050, Romania}

\correspondence{Correspondence should be addressed to 
    	Mehedi Kalam: kalam@associates.iucaa.in}

\email{masummurshid2012@gmail.com (M. Murshid), rahmannilofar@gmail.com (N. Rahman), radinschi@yahoo.com (I. Radinschi)}


\maketitle 

\begin{abstract}
The low mass compact stars are quite fascinating objects to study for their enigmatic behaviour. In this paper, we have modeled this kind of low mass strange stars based on the Heintzmann ansatz \cite{Heintzmann1969} in $(2+1)$ dimension. Attractive anisotropic force plays a significant role to restrict the upper mass limit (which is comparatively low) of the strange star. We have applied our model to some low mass strange stars. Our model could be useful to predict the important parameters of the low mass strange stars.

\end{abstract}
    
\section{Introduction}
The peculiar behaviour of some compact stars opens the door of the
possibility of the low mass neutron stars and the low mass strange stars.
One can explain the high braking index of the PSR J1640-4631 \cite{Chen2016}
and the smaller polar cap area of the PSR B0943+10 \cite{Yue2006} by
considering that we can say these are low mass strange stars. Though
theoretically, the low mass compact stars could be a product of a
core-collapse supernova, it is in reality very unlikely that low mass
compact stars are created from the supernova. These low mass compact stars
are formed as the massive white dwarf collapses due to accretion \cite{Xu2005,Xu2006}. These compact stars could be either a neutron star or
strange star. There are few ways of distinguishing a neutron star and a
strange star. The Mass-radius relation of the star is one of the ways of
distinguishing the neutron star and strange star. Mass of neutron star, $
M_{ns} \propto R^{-3}$ , whereas mass of strange star, $M_{ss}\propto R^{3}$ 
\cite{Alcock1986,Bombaci1997,Li1999,Xu2005}. A neutron star having mass $
\sim 0.2 M_{\odot}$ has a radius of $> 15 km $, whereas a strange star with
mass $\sim 0.2 M_{\odot}$ is only $< 5 km $. So, we see that for low mass
neutron and strange star with the same mass show a significant difference in
radius whereas a neutron star and strange star with $\sim 1 M_{\odot}$ mass
and above have almost the same radii \cite{Xu2005}.
\par Recently lower-dimensional gravity has become valuable due to its simplicity
in describing the geometry of spacetime. It has illuminated the haziness
surrounding the four-dimensional gravity. Banados, Teitelboim and Zanelli
(BTZ) described the (2+1) dimensional spacetime geometry with a negative
cosmological constant and which admits a black hole solution \cite{Banados1992}. This work was revolutionary back then. It is easier to deal
with a set of not so complicated equations since the system imitates the
four-dimensional analysis. It is fascinating that BTZ black hole is a
solution of low energy string theory with a non-vanishing antisymmetric
tensor and it resembles with the exterior of a (2+1) dimensional perfect
fluid star. Keeping this in mind, Cruz and Zanelli \cite{Cruz1995} obtained
the interior solution putting upper limit for mass regarding the generic
equation of state $P=P(\rho)$. By a simple dimensional reduction, it is
possible to get a (2+1) dimensional perfect fluid solution with constant
energy density which can be obtained from the Schwarzschild interior metric
by comparing (2+1) and (3+1) gravity- it was an important interpretation
given by Garcia and Campuzano \cite{Garcia2003}. Mann and Ross \cite{Mann1993} analysed that it is possible for a (2+1) dimensional star which
is filled with dust ($\rho=0$) to collapse to a black hole under some
certain conditions. There is an exact solution in (2+1) dimensional gravity
with a negative cosmological constant, for the critical collapse of a scalar
field in the closed-form given by Garfinkle \cite{Garfinkle2001}. Sa \cite{Sa1999} also gave another solution assuming a polytropic equation of state
of the form $P=K \rho^{1+\frac{1}{n}}$, where `$K$' is polytropic constant
and `$n$' is the polytropic index. Sharma et al. \cite{Sharma2011} have also
taken a particular form of the mass function to study the interior of an
isotropic star in (2+1) dimensional gravity. On the other hand, Rahaman et
al. \cite{Rahaman2013} and Bhar et al. \cite{Bhar2015} have separately
studied non-singular model for anisotropic stars based on the Krori and
Barua (KB) ansatz in (2 + 1) dimension. Some researcher \cite{Banerjee2013,Bhar2014} has presented a class of interior solutions corresponding to the BTZ exterior by using Finch and Skea ansatz.

The purpose of the present work is to construct a low mass strange star
model based on the Heintzmann ansatz in $(2+1)$ dimensions. The motivation
for doing so is due to the curiosity of the role of anisotropy to bound the
upper mass limit (which is comparatively low) of the strange star. The plan
of this paper is as follows. In Sec. 2, we discuss the interior spacetime of
the low mass strange stars. In Sec. 3, we look at some physical properties
of the strange star. In sub-section, we discuss the Matching condition with
exterior BTZ solution, Behaviour of energy density and anisotropic pressure,
compactness, surface redshift, energy condition and validity of generalised
TOV equation. In Sec. 4, we discuss several conditions imposed on the metric
parameters. In Sec. 5, we apply our model on the three different compact
objects. We discuss our results in Sec. 6.

\section{Interior Spacetime}

The line element which describes the interior spacetime of a static
spherically symmetric compact object in $(2+1)$ dimension is written as 
\begin{equation}
ds^{2}=-e^{2\nu (r)}dt^{2}+e^{2\mu (r)}dr^{2}+r^{2}d\phi ^{2}.  \label{eq:1}
\end{equation}

The energy-momentum tensor for the matter distribution in the interior of
the anisotropic star has the standard form as 
\begin{equation}
T_{ij}=diag(-\rho ,p_{r},p_{t}),  \label{eq:2}
\end{equation}
where $\rho $, $p_{r}$ and $p_{t}$ represent the energy density, normal
radial pressure and transverse pressure respectively. 

The Einstein's equations for the metric with negative cosmological constant $
(\Lambda <0)$ in geometric unit $(G=c=1)$ can be written as 
\begin{eqnarray}
2\pi \rho +\Lambda  &=&\frac{ e^{ -2\mu }\mu^{\prime}}{r},  \label{eq:3} \\
2\pi p_{r}-\Lambda  &=&\frac{ e^{ -2\mu }\nu ^{\prime}}{r},  \label{eq:4} \\
2\pi p_{t}-\Lambda  &=&e^{-2\mu }(\nu ^{\prime \prime }+\nu ^{\prime 2}-\nu
^{\prime }\mu ^{\prime }).  \label{eq:5}
\end{eqnarray}
From eqn.(3) we can get the radial dependent mass function (taking
integration const. as unity\cite{Sharma2011}) as 
\begin{equation}
m(r)=\int_{0}^{r}2\pi \rho \tilde{r}d\tilde{r}=\frac{1}{2}(1-e^{-2\mu
}-\Lambda r^{2}).  \label{eq:6}
\end{equation}
In 1916 Schwarzschild  \cite{Schwarzschild1916} first solved the exact solution of Einstein's field equations; and Oppenheimer, Volkoff and Tolman \cite{Oppenheimer1939, Tolman1939} in 1939 successfully derived the balancing equations of relativistic stellar structures from Einstein's field equations. Since then, several scientists trying to get a new exact solution of Einstein's field equations for the interior region of the stars and unfolding several new aspects of nature. Recently, some scientists \cite{Maurya:2015maa,Maurya:2018kxg,Maurya:2017uyk,Gupta:2010bym,Maurya:2019kzu,MAURYA2012677,Maurya:2019one,Dev:2000gt,2014IJTP...53.3958P}, In this paper, we use Heintzmann's exact solution in (2+1) dimensions to explore some new features of the compact stars. According to Heintzmann \cite{Heintzmann1969} 
\begin{eqnarray}
e^{2\nu } &=&A^{2}(1+ar^{2})^{3},  \label{eq:7} \\
e^{-2\mu } &=&1-\frac{3ar^{2}}{2}\left[ \frac{1+C(1+4ar^{2})^{-\frac{1}{2}}}{%
1+ar^{2}}\right] .  \label{eq:8}
\end{eqnarray}
where $A$ and $C$ are dimensionless constant and $a$ is a constant with
dimension of $length^{-2}$ in geometric unit.

Therefore, the mass function comes out as

\begin{equation}
m(r)=\frac{3ar^{2}\left[ 1+C\left( 1+4ar^{2}\right) ^{-\frac{1}{2}}\right] }{
4\left( 1+ar^{2}\right) }-\frac{\Lambda r^{2}}{2},  \label{eq:9}
\end{equation}

which is regular at the centre i.e. $m(r)=0$ at $r=0$.

Solving from eqn.(\ref{eq:3}-\ref{eq:8}) we get 
\begin{equation}
\rho =\frac{3a\left[ 2aCr^{2}\left( 1-ar^{2}\right) +\left( 4ar^{2}+1\right)
^{3/2}+C\right] }{4\pi \left( ar^{2}+1\right) ^{2}\left( 4ar^{2}+1\right)
^{3/2}}-\frac{\Lambda }{2\pi },  \label{eq:10}
\end{equation}
\begin{small}
\begin{equation}
p_{r}=\frac{3a\left[ -ar^{2}\left\{ 3C\left( 4ar^{2}+1\right) ^{-\frac{1}{2}
}+1\right\} +2\right] }{4\pi \left( ar^{2}+1\right) ^{2}}+\frac{\Lambda }{
2\pi },  \label{eq:11}
\end{equation}
\begin{eqnarray}
&&p_{t}=\frac{3a}{2\pi \left( ar^{2}+1\right) ^{2}\left( 4ar^{2}+1\right) } 
\nonumber  \label{eq:12} \\
&&\times \left[ 1-ar^{2}\left\{ 3C\frac{\left( 3ar^{2}+1\right) }{\left(
4ar^{2}+1\right) ^{\frac{1}{2}}}+\left( 4ar^{2}-3\right) \right\} \right] +
\frac{\Lambda }{2\pi }.  \nonumber \\
&&
\end{eqnarray}
\end{small}

Then the central density and pressure are 
\begin{eqnarray}
\rho _{0} &=&\frac{3a(C+1)-2\Lambda }{4\pi },  \label{eq:13} \\
p_{0} &=&\frac{3a+\Lambda }{2\pi }.  \label{eq:14}
\end{eqnarray}
Thus, central density and pressure should remain positive provided $a>\frac{
\mid \Lambda \mid }{3}$ where $a$ and $C$ should be positive.

\begin{figure*}[!htbp]
\centering
\includegraphics[width=1\textwidth]{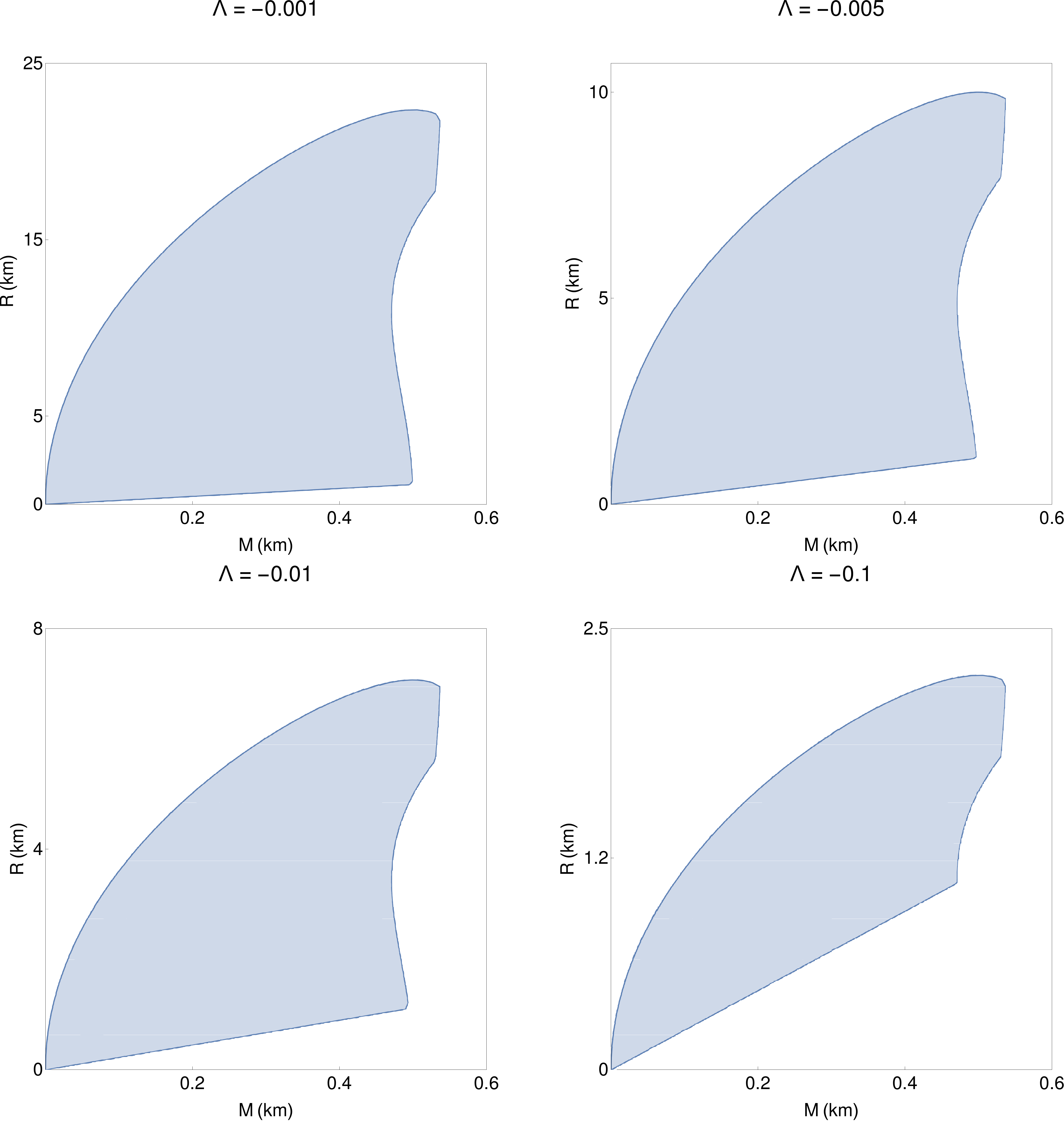}
\caption{Graphical presentation of accessible mass radius region of our
model }
\label{fig:1}
\end{figure*}

\begin{table*}[!ht]
\begin{tabular}{@{}p{5.4cm}p{5.4cm}p{5.4cm}}
\hline
Cosmological Constant $(\Lambda)$ & Mass $(M)$ in $km$ & Radius $(R)$ in $km$
\\ \hline
$\Lambda \leq -0.395062$ & $0<M<M_{1}$ & $2.25M<R<R_{1}$ \\[3ex] \hline
$-0.395062<\Lambda \leq -0.332447$ & $0<M<0.5$ & $2.25M<R<R_{1} $ \\[1ex] 
& $0.5\leq M<M_{1} $ & $2.25M<R<R_{1} $ \\[3ex] \hline
& $0<M<0.5 $ & $2.25M<R<R_{1} $ \\[1ex] 
$-0.332447<\Lambda \leq -0.221879$ & $0.5\leq M\leq M_{2} $ & $2.25M<R<R_{1} 
$ \\[1ex] 
& $M_{2}<M<0.536438 $ & $R_{2}<R<R_{1}$ \\[3ex] \hline
& $0 < M < 0.5 $ & $2.25M < R <R_{1}$ \\[1ex] 
& $0.5\leq M \leq M_{3}$ & $2.25M < R <R_{1}$ \\[1ex] 
$-0.221879<\Lambda<-0.197531$ & $M_{3}<M<0.530149$ & $R_{3}\leq R <R_{1}$ \\
[1ex] 
& $M=0.530149$ & $\frac{0.561874}{\sqrt{|\Lambda|}}< R<\frac{0.700116}{\sqrt{
|\Lambda|}}$ \\[1ex] 
& $0.530149<M<0.536438$ & $R_{2}<R< R_{1}$ \\[1ex] \hline
& $0 < M < 0.5$ & $2.25 M < R <R_{4}$ \\[1ex] 
& $M =0.5$ & $1.125 < R < 1.59099$ \\[1ex] 
$\Lambda = -0.197531 $ & $0.5 < M < 0.530149 $ & $R_{4}\leq R<R_{5}$ \\[1ex] 
& $M = 0.530149$ & $1.26422 < R < 1.57526$ \\[1ex] 
& $0.530149 <M<0.536438$ & $R_{6}<R<R_{5}$ \\[1ex] \hline
& $0<M<M_{3}$ & $2.25M<R<R_{1}$ \\[1ex] 
& $M_{3}<M<0.5$ & $R_{3}\leq R <R_{1}$ \\[1ex] 
$-0.197531 < \Lambda \leq -0.104938$ & $M=0.5$ & $\frac{1}{2\sqrt{|\Lambda|}}
\leq R < \frac{1}{\sqrt{2|\Lambda|}}$ \\[1ex] 
& $0.5 < M < 0.530149$ & $R_{3}\leq R<R_{1}$ \\[1ex] 
& $M =0.530149 $ & $\frac{0.561874}{\sqrt{|\Lambda|}}< R<\frac{0.700116}{
\sqrt{|\Lambda|}}$ \\[1ex] 
& $0.530149 < M < 0.536438$ & $R_{2}< R<R_{1}$ \\[1ex] \hline
& $0 < M \leq 0.470588$ & $2.25M<R<R_{1}$ \\[1ex] 
& $0.470588 < M <M_{3}$ & $2.25 M < R\leq R_{7}$ or $R_{3}\leq R <R_{1}$ \\
[1ex] 
& $M_{3}\leq M<0.5$ & $R_{3}\leq R <R_{1} $ \\[1ex] 
$-0.104938 < \Lambda < 0$ & $M=0.5$ & $\frac{1}{2\sqrt{|\Lambda|}} \leq R < 
\frac{1}{\sqrt{2|\Lambda|}}$ \\[1ex] 
& $0.5 < M < 0.530149$ & $R_{3}\leq R<R_{1} $ \\[1ex] 
& $M = 0.530149$ & $\frac{0.561874}{\sqrt{|\Lambda|}}< R<\frac{0.700116}{
\sqrt{|\Lambda|}}$ \\[1ex] 
& $0.530149 < M < 0.536438$ & $R_{2}\leq R<R_{1} $ \\[1ex] \hline
\end{tabular}
\caption{List of the possible range of the radius and mass of the star and
cosmological constant where $M_{2}$, $M_{3}$, $R_{2}$ and $R_{6}$ are the
real positive root of $M_{2}(x)$, $M_{3}(x)$, $R_{2}(x)$ and $R_{6}(x)$
respectively ($M_{2}(x)$, $M_{3}(x)$, $R_{2}(x)$, $R_{6}(x)$ , $R_{1}$, $
R_{3}$, $R_{4}$, $R_{5}$ and $R_{7}$ are given in Appendix:A )}
\label{tab:app2}
\end{table*}

\begin{table}[!htbp]
\begin{tabular}{ccc}
\hline
$\Lambda$ & $M$ in $km$ & $R$ in $km$ \\ \hline
& 0.001 & $0.00225 < R < 1.15457$ \\ 
& 0.005 & $0.01125 < R < 2.58055$ \\ 
& 0.01 & $0.0225 < R < 3.64739$ \\ 
-0.001 & 0.05 & $0.1125 < R < 8.11733$ \\ 
& 0.1 & $0.225 < R < 11.405$ \\ 
& 0.5 & $15.8114 \leq R < 22.3607$ \\ 
& 0.53 & $17.7598 \leq R < 22.1422$ \\[2ex] 
& 0.001 & $0.00225 < R < 0.51634$ \\ 
& 0.005 & $0.01125 < R < 1.15406$ \\ 
& 0.01 & $0.0225 < R < 1.63116$ \\ 
-0.005 & 0.05 & $0.1125 < R < 3.63018$ \\ 
& 0.1 & $0.225 < R < 5.10046$ \\ 
& 0.5 & $7.07107 \leq R < 10.0$ \\ 
& 0.53 & $7.94242 \leq R < 9.9023$ \\[2ex] 
& 0.001 & $0.00225 < R < 0.365108$ \\ 
& 0.005 & $0.01125 < R < 0.816041$ \\ 
& 0.01 & $0.0225 < R < 1.15341$ \\ 
-0.01 & 0.05 & $0.1125 < R < 2.56693$ \\ 
& 0.1 & $0.225 < R < 3.60657$ \\ 
& 0.5 & $5.0 \leq R < 7.07107$ \\ 
& 0.53 & $5.61614\leq R < 7.00198$ \\[2ex] 
& 0.001 & $0.00225 < R < 0.115457$ \\ 
& 0.005 & $0.01125 < R < 0.258055$ \\ 
& 0.01 & $0.0225 < R < 0.364739$ \\ 
-0.1 & 0.05 & $0.1125 < R < 0.811733$ \\ 
& 0.1 & $0.225 < R < 1.1405$ \\ 
& 0.5 & $1.58114 \leq R < 2.23607$ \\ 
& 0.53 & $1.77598 \leq R < 2.21422$ \\ \hline
\end{tabular}
\caption{List of the numerical values the cosmological constant $\Lambda$,
mass $M$ and radius $R$ in the view of our model }
\label{tab:app1}
\end{table}

\begin{figure*}[!b]
\includegraphics[width=1\textwidth]{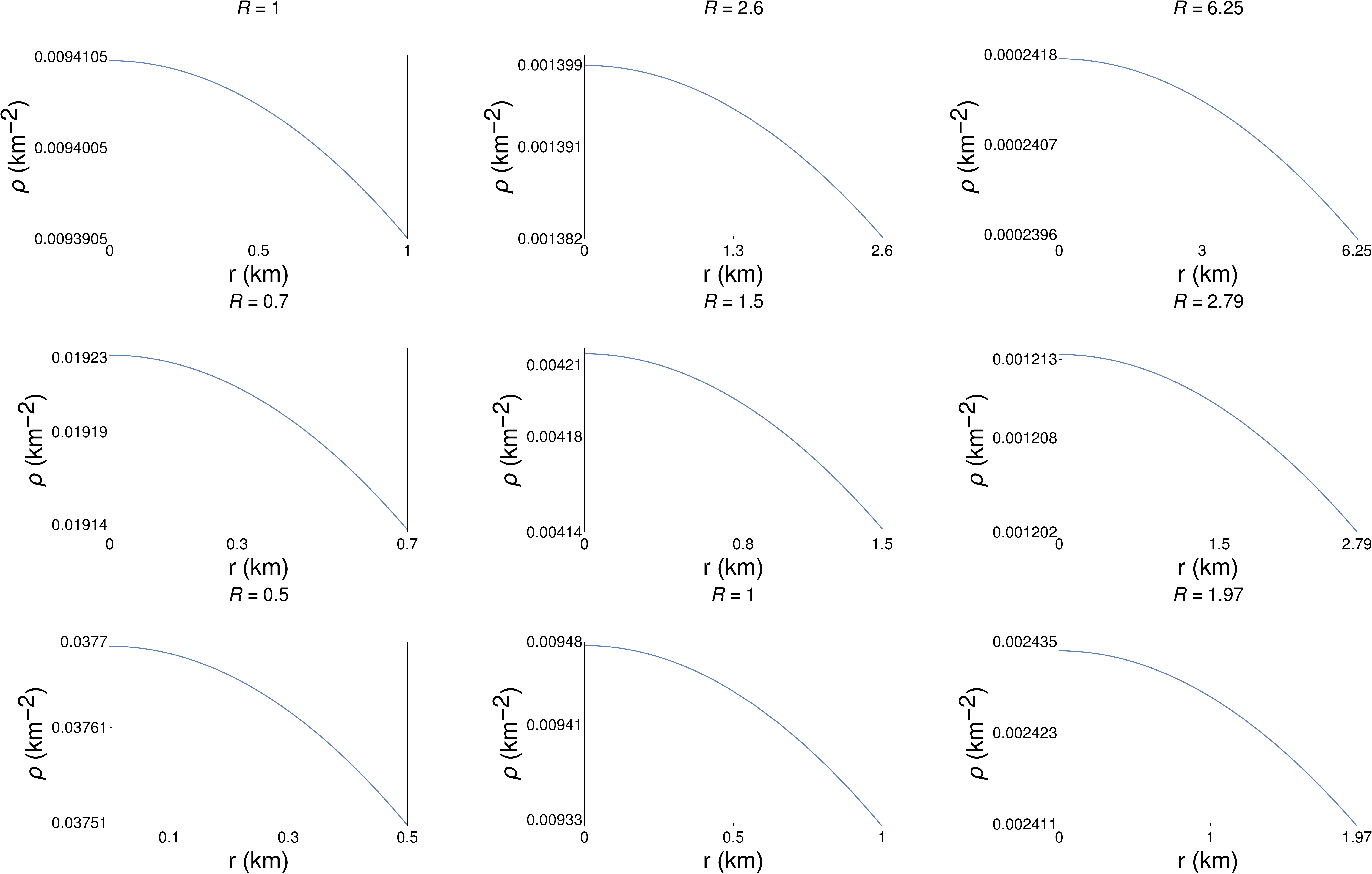}
\caption{The variation of the energy density with radial coordinate for the
Pulsar PSR B0943+10 (1st row for $\Lambda=-0.001 km^{-2}$ , 2nd row for $
\Lambda=-0.005 km^{-2}$ and 3rd row for $\Lambda=-0.01 km^{-2}$)}
\label{fig:2}
\end{figure*}

\begin{figure*}[!htbp]
\includegraphics[width=1\textwidth]{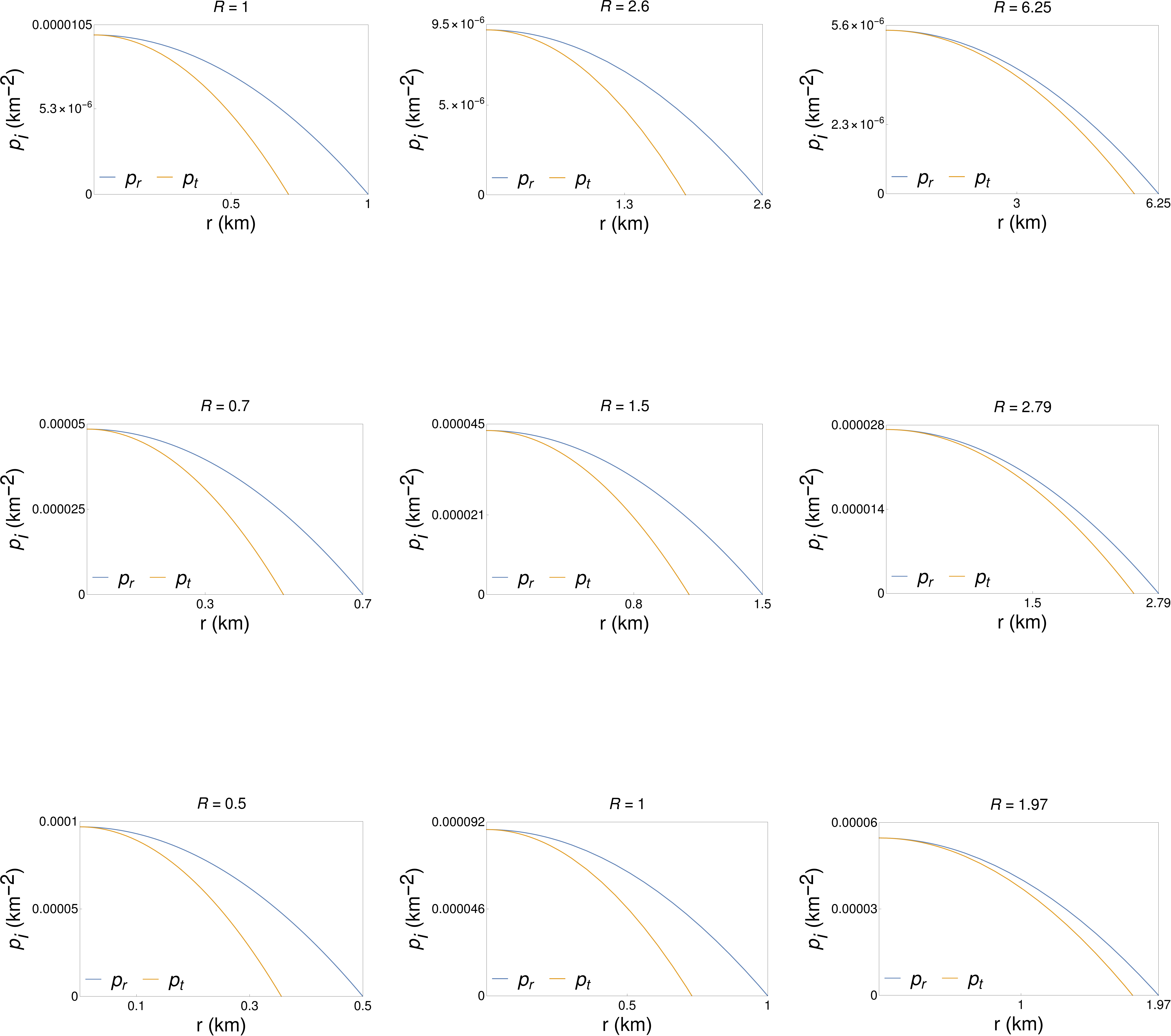}
\caption{The variation of the radial and transverse pressure with radial
coordinate for the Pulsar PSR B0943+10 (1st row for $\Lambda=-0.001 km^{-2}$
, 2nd row for $\Lambda=-0.005 km^{-2}$ and 3rd row for $\Lambda=-0.01 km^{-2}
$) }
\label{fig:3}
\end{figure*}

\begin{figure*}[!htbp]
\includegraphics[width=1\textwidth]{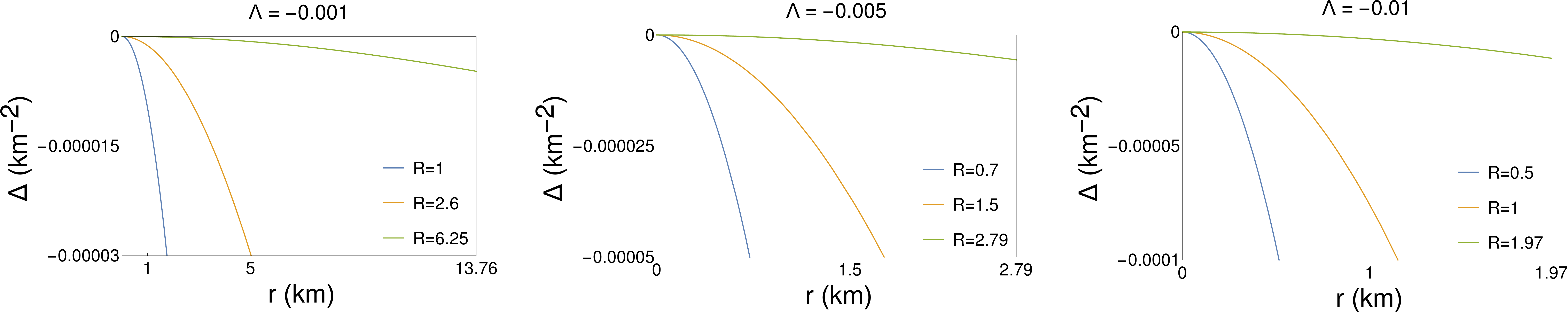}
\caption{The variation of the anisotropic force with radial coordinate for
the Pulsar PSR B0943+10}
\label{fig:4}
\end{figure*}

\begin{figure*}[!htbp]
\includegraphics[width=1\textwidth]{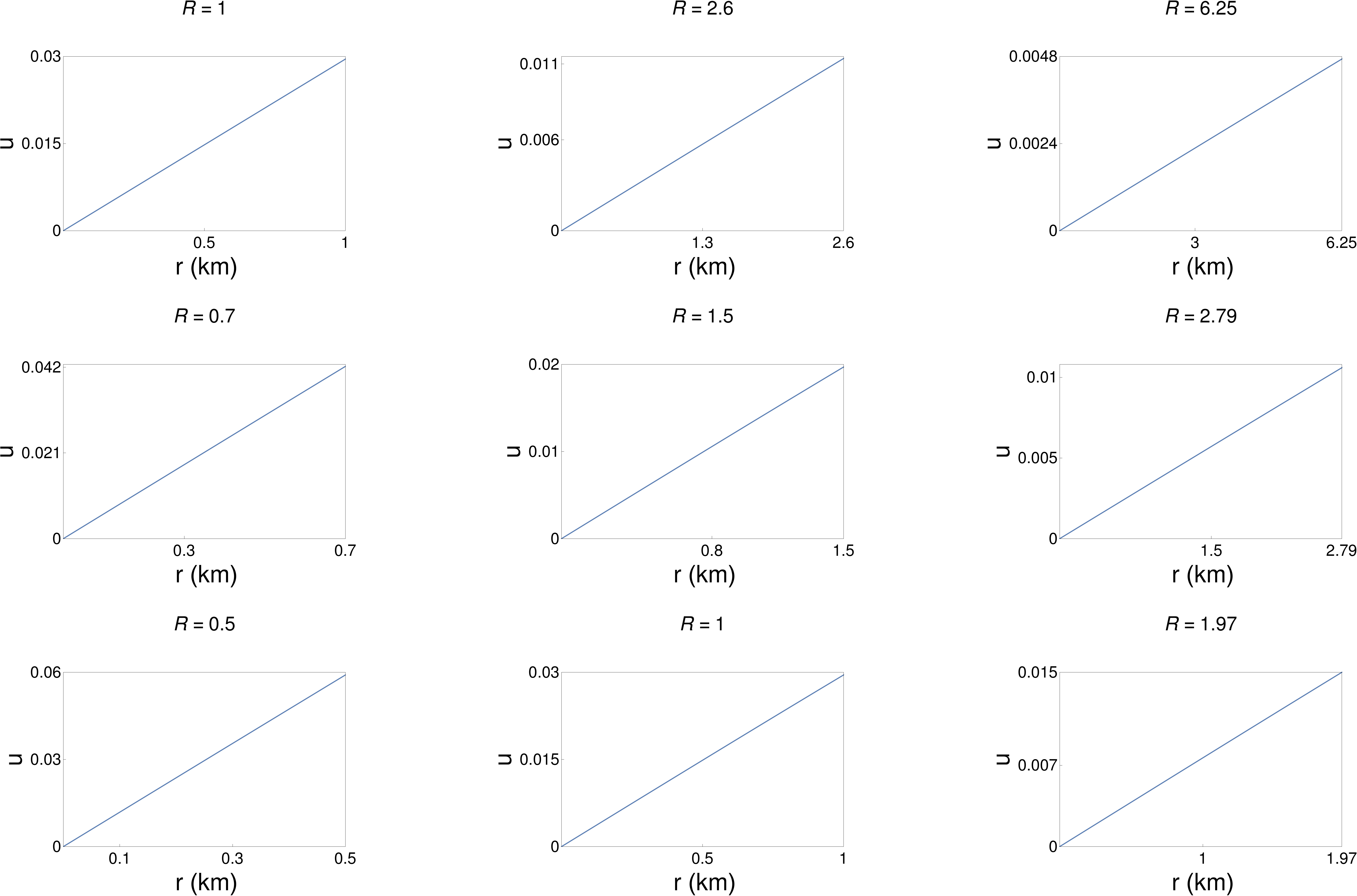}
\caption{The variation of the compactness with radial coordinate for the
Pulsar PSR B0943+10 (1st row for $\Lambda=-0.001 km^{-2}$ , 2nd row for $
\Lambda=-0.005 km^{-2}$ and 3rd row for $\Lambda=-0.01 km^{-2}$)}
\label{fig:5}
\end{figure*}

\begin{figure*}[!htbp]
\includegraphics[width=1\textwidth]{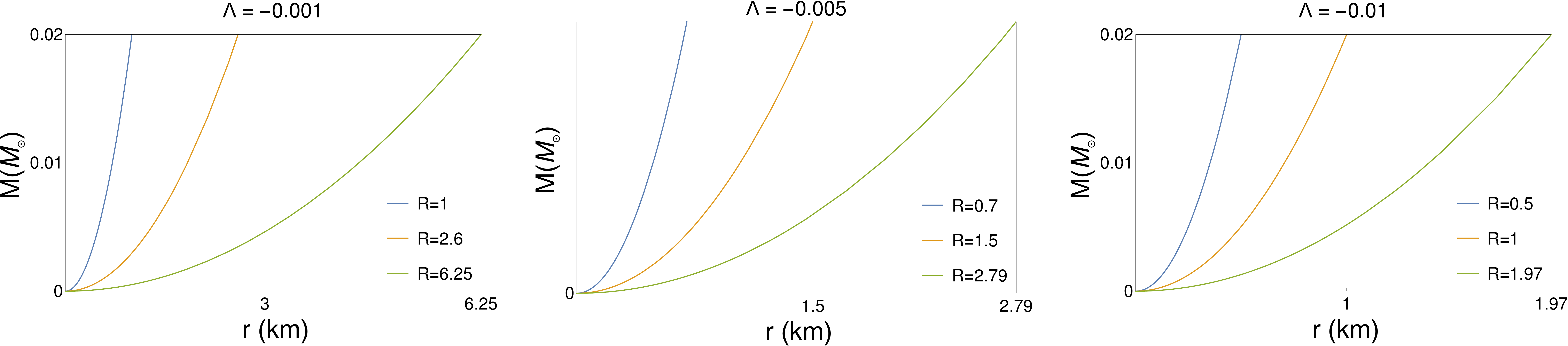}
\caption{The variation of mass function is shown for the Pulsar PSR B0943+10 
}
\label{fig:6}
\end{figure*}

\begin{figure*}[!htbp]
\includegraphics[width=1\textwidth]{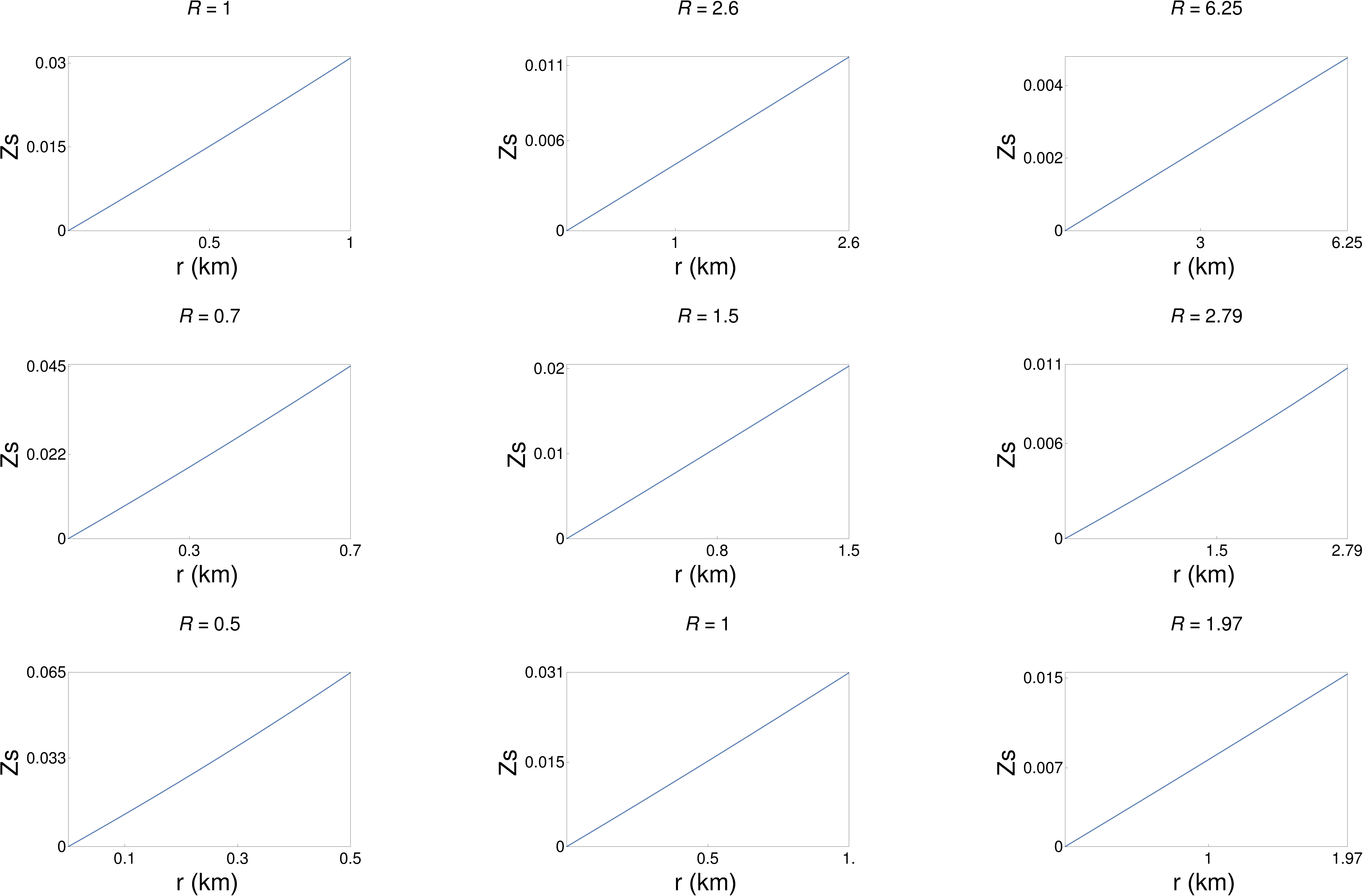}
\caption{The variation of the redshift with radial coordinate for the Pulsar
PSR B0943+10 (1st row for $\Lambda=-0.001 km^{-2}$ , 2nd row for $
\Lambda=-0.005 km^{-2}$ and 3rd row for $\Lambda=-0.01 km^{-2}$) }
\label{fig:7}
\end{figure*}

\begin{figure*}[!htbp]
\includegraphics[width=1.1\textwidth]{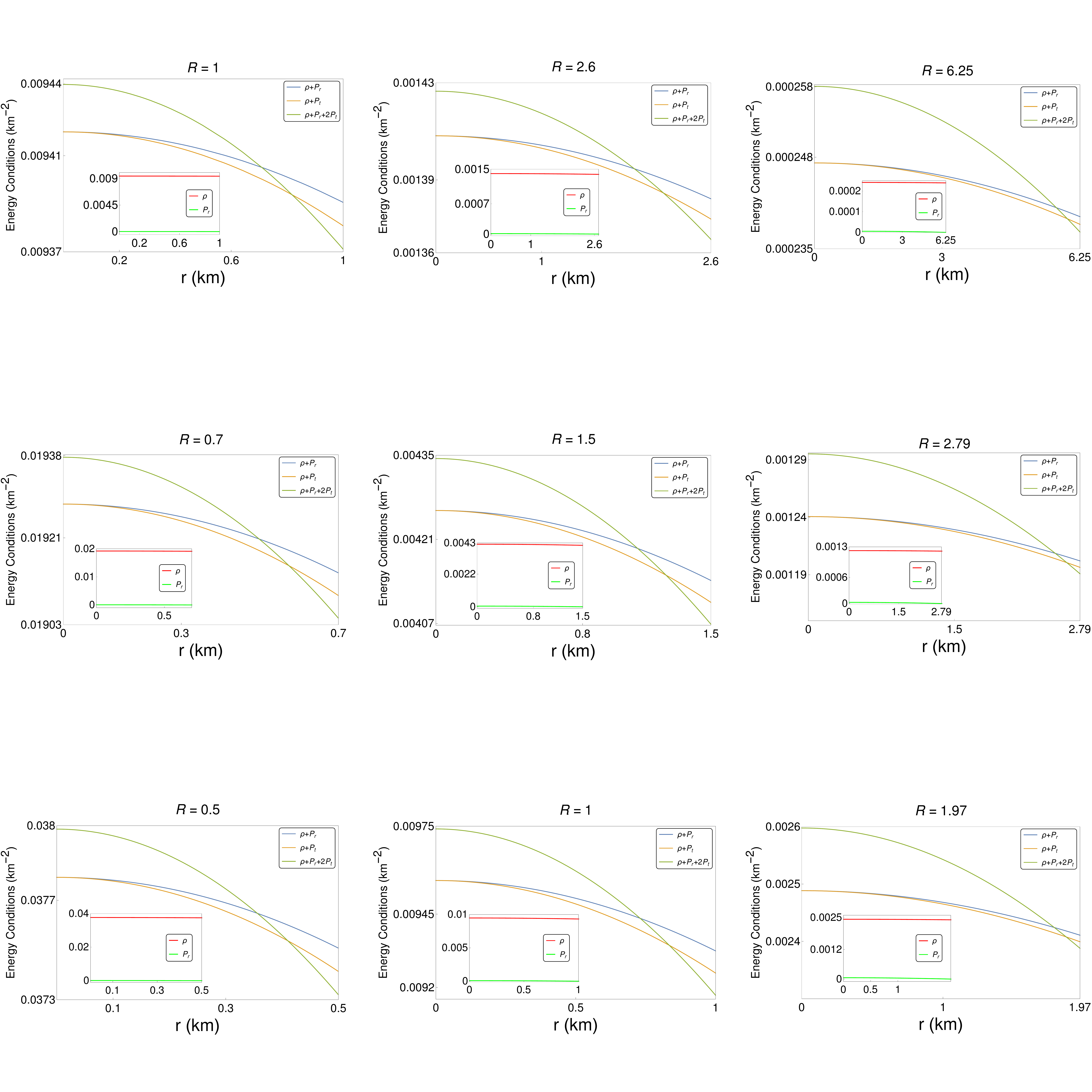}
\caption{Energy Conditions at the interior of the star for the Pulsar PSR
B0943+10 (1st row for $\Lambda=-0.001 km^{-2}$ , 2nd row for $\Lambda=-0.005
km^{-2}$ and 3rd row for $\Lambda=-0.01 km^{-2}$) }
\label{fig:8}
\end{figure*}

\begin{figure*}[!htbp]
\includegraphics[width=1\textwidth]{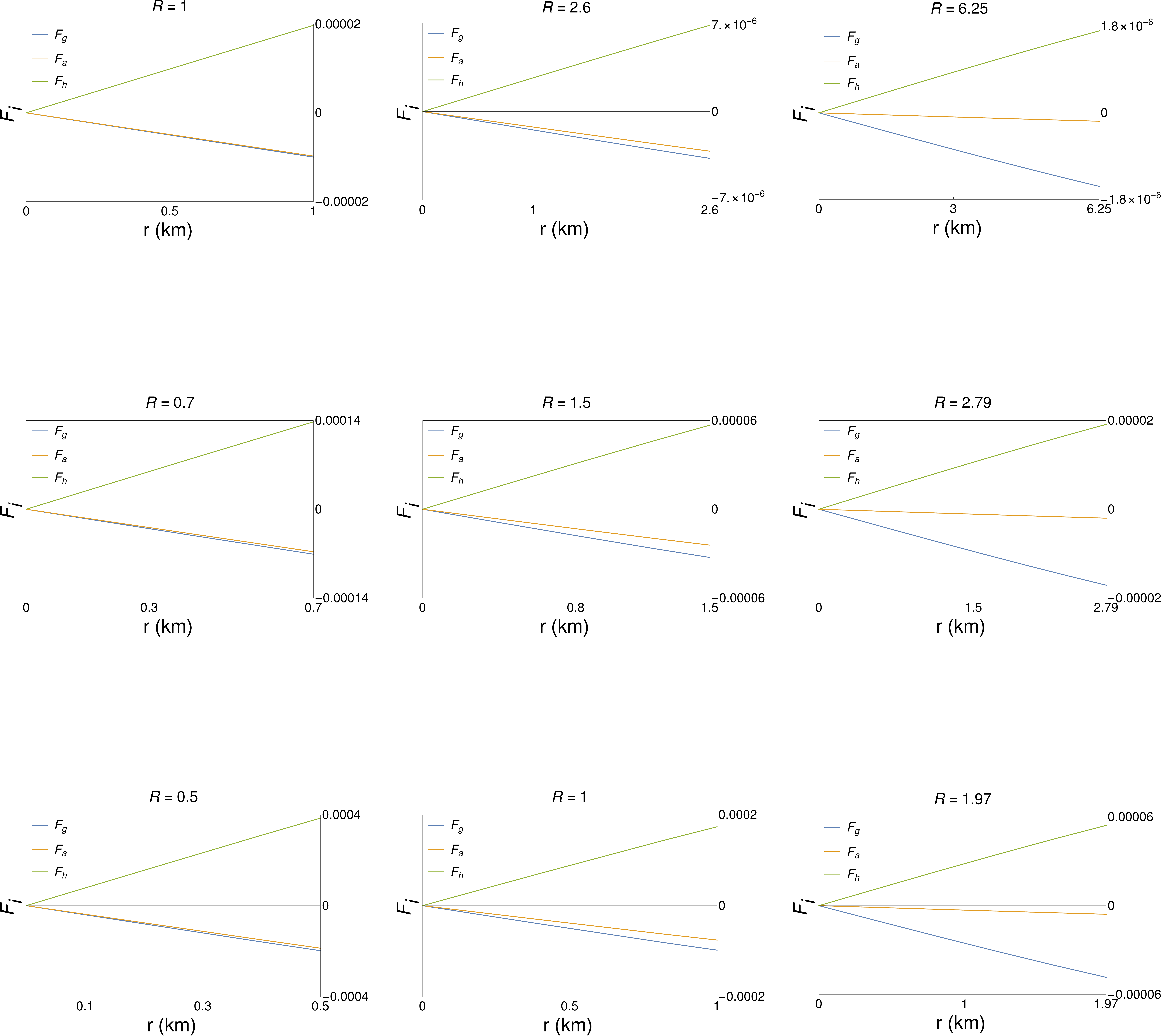}
\caption{The behavior of three different forces acting on the fluid is shown
for the Pulsar PSR B0943+10 (1st row for $\Lambda=-0.001 km^{-2}$ , 2nd row
for $\Lambda=-0.005 km^{-2}$ and 3rd row for $\Lambda=-0.01 km^{-2}$) }
\label{fig:9}
\end{figure*}

\begin{figure}[!htbp]
\includegraphics[width=0.48\textwidth]{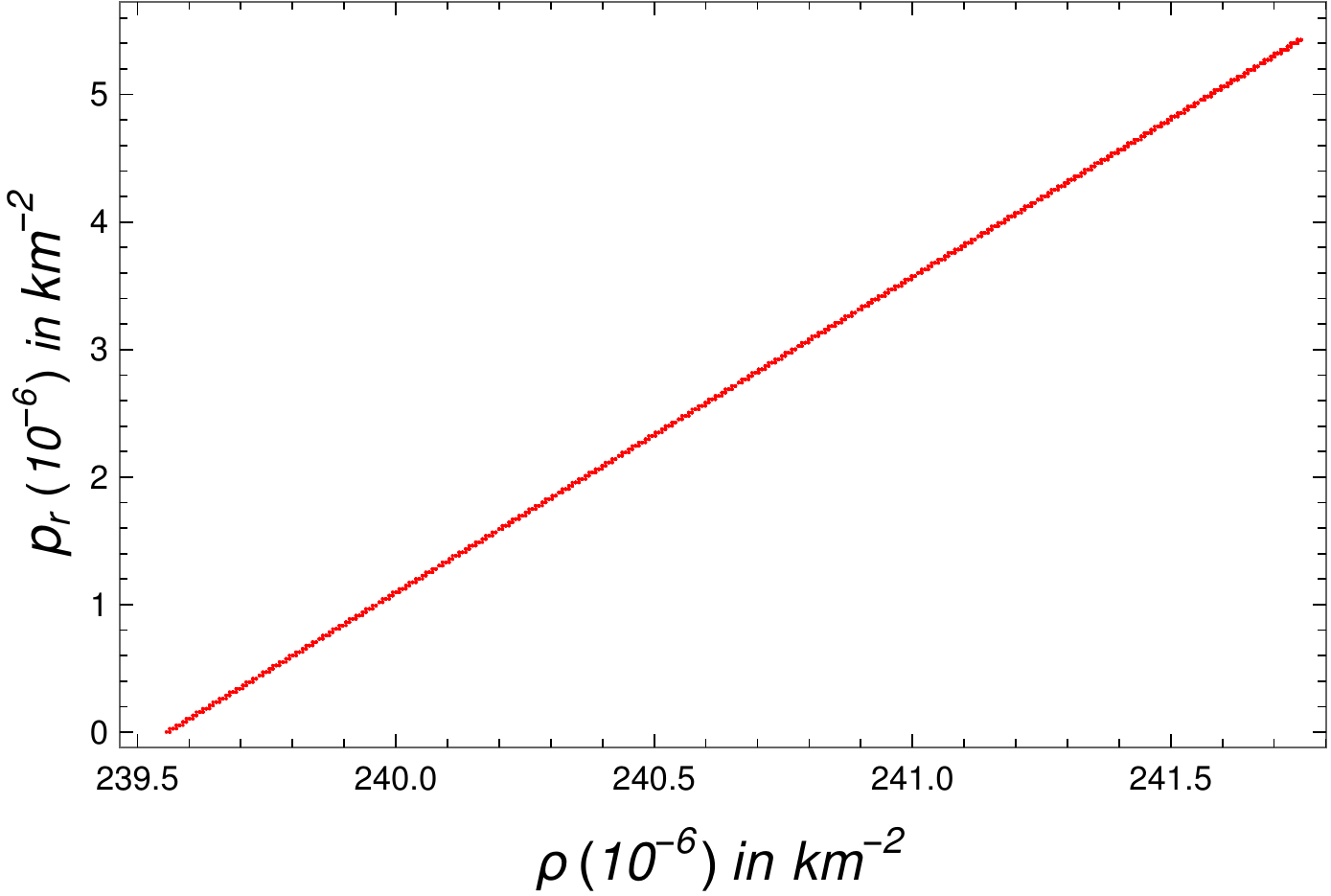}
\caption{The relation between radial pressure ($p_{r}$) and energy density ($\rho$) (EoS)in the stellar interior region (Taking $M=0.02M_{\odot}$, $R=6.25km$ and $\Lambda=-0.001 km^{-2}$)}
\label{fig:10}
\end{figure}

\section{Some Physical Properties}

\subsection{Matching condition}

The exterior spacetime of the static spherically symmetric compact object is
assumed to be described by the BTZ metric as follow 
\begin{equation}
ds^{2}=-(-M_{0}-\Lambda r^{2})dt^{2}+\frac{dr^{2}}{-M_{0}-\Lambda r^{2}}
+r^{2}d\phi ^{2}.  \label{eq:15}
\end{equation}
 \par The parameter $M_{0}$ is the conserved charge associated with asymptotic invariance under the time displacements.
The continuity of $g_{tt}$ and $g_{rr}$ at surface $(r=R)$ and vanishing of
normal pressure at surface yield 
\begin{eqnarray}
e^{2\nu (R)} &=&-M_{0}-\Lambda R^{2},  \label{eq:16} \\
e^{-2\mu (R)} &=&-M_{0}-\Lambda R^{2},  \label{eq:17} \\
0 &=&\frac{\Lambda }{2\pi }+\frac{\nu ^{\prime -2\mu (R)}}{2\pi R}.
\label{eq:18}
\end{eqnarray}

Solving these above three equations (eqn.-\ref{eq:16}-\ref{eq:18}) we get 
\begin{equation}
A=\frac{1}{3\sqrt{3}}\sqrt{-\frac{\left( 3M_{0}+2\Lambda R^{2}\right) {}^{3}
}{\left( M_{0}+\Lambda R^{2}\right) {}^{2}},}  \label{eq:19}
\end{equation}
\begin{equation}
a=\frac{\Lambda }{3M_{0}+2\Lambda R^{2}},  \label{eq:20}
\end{equation}
\begin{small}
\begin{eqnarray}
C &=&\frac{\sqrt{3}}{\Lambda R^{2}}\sqrt{\frac{M_{0}+2\Lambda R^{2}}{
3M_{0}+2\Lambda R^{2}}}  \nonumber  \label{eq:21} \\
&&\times \left[ M_{0}\left( 4\Lambda R^{2}+2\right) +2M_{0}^{2}+\Lambda
R^{2}\left( 2\Lambda R^{2}+1\right) \right] \nonumber \\
\end{eqnarray}
\end{small}
The total mass of the compact object of radius $R$ is given by 
\begin{equation}
M(R)=\frac{1}{2}(1-e^{-2\mu (R)}-\Lambda R^{2})=\frac{1}{2}(1+M_{0}).
\label{eq:22}
\end{equation}

\subsection{Behavior of energy density and pressure}

For a physically acceptable model, energy density and radial pressure both
should be monotonically decreasing function in $r$ and should be maximum at
the centre. For our model, we have

 \begin{small}
\begin{eqnarray}
\frac{d\rho }{dr} &=&-\frac{3a^{2}r}{\pi \left( ar^{2}+1\right) ^{3}\left(
4ar^{2}+1\right) ^{5/2}}  \nonumber  \label{eq:23} \\
&&\times \left[ C\left( -6a^{3}r^{6}+12a^{2}r^{4}+12ar^{2}+3\right) +\left(
4ar^{2}+1\right) ^{5/2}\right] <0  \nonumber \\
,  \nonumber \\
&&
\end{eqnarray}
\begin{eqnarray}
&&\frac{dp_{r}}{dr}=-\frac{3a^{2}r}{2\pi \left( ar^{2}+1\right) ^{3}\left(
4ar^{2}+1\right) ^{3/2}}  \nonumber  \label{eq:24} \\
&&\times \left[ 3C\left( 6a^{2}r^{4}-ar^{2}-1\right) \right.   \nonumber \\
&&+\left. \sqrt{4ar^{2}+1}\left( 4a^{2}r^{4}-19ar^{2}-5\right) \right] <0.
\end{eqnarray}
\end{small}

At the centre ($r=0$), 
\begin{equation}
\frac{d\rho }{dr}\bigg|_{r=0}=\frac{dp_{r}}{dr}\bigg|_{r=0}=0  \label{eq:25}
\end{equation}
and 
\begin{eqnarray}
\frac{d^{2}\rho }{dr^{2}}\bigg|_{r=0} &=&-\frac{3a^{2}(3C+1)}{\pi }<0,
\label{eq:26} \\
\frac{d^{2}p_{r}}{dr^{2}}\bigg|_{r=0} &=&-\frac{3a^{2}(3C+5)}{2\pi }<0.
\label{eq:27}
\end{eqnarray}

The above eqn.(\ref{eq:25}-\ref{eq:27}) imply that central energy density
and central pressure are maximum at the centre for any positive value of $a$
and $C$.

The measure of anisotropy for our model is given by the expression 
\begin{equation}
\Delta =p{_{t}}-p_{r}=-\frac{3a^{2}r^{2}\left[ C\left( 6ar^{2}+3\right)
+\left( 4ar^{2}+1\right) ^{3/2}\right] }{4\pi \left( ar^{2}+1\right)
^{2}\left( 4ar^{2}+1\right) ^{3/2}}.  \label{eq:28}
\end{equation}

Based on the sign of the anisotropy parameter, the anisotropic force can be
categorised in two: (i) the repulsive anisotropic force when $p_t > p_r$
i.e. $\Delta>0$ and (ii) the attractive anisotropic force when $p_t < p_r$
i.e. $\Delta<0$ \cite{Kalam2012}. This repulsiveness in anisotropic force enhances the stability of the star resulting in the star to be more compact than the isotropic one.
\cite{Gokhroo:1994fbj,Roupas:2020mvs,Molla:2020eib}

\subsection{Energy conditions}

The energy conditions such as null energy condition (NEC), weak energy
condition (WEC), strong energy condition (SEC) and dominant energy condition
(DEC) should be satisfied at every point in the interior of the compact star
simultaneously. These energy conditions are as follows: \newline

i) NEC: $\rho+p_{i} \geq 0 $ ;\newline
ii) WEC: $\rho+p_{i} \geq 0 $ , $\rho \geq 0 $ ;\newline
iii) SEC: $\rho+p_{i} \geq 0$ , $\rho+p_{r}+2p_{t} \geq 0 $ ;\newline
iv) DEC: $\rho >|p_{i}| $ ; \newline

\subsection{Compactness and Surface Redshift}

The compactness of the star is given by 
\begin{equation}
u=\frac{m(r)}{r}=\frac{3ar\left[ 1+C\left( 1+4ar^{2}\right) ^{-\frac{1}{2}}
\right] }{4\left( 1+ar^{2}\right) }-\frac{r\Lambda }{2}.  \label{eq:29}
\end{equation}

The corresponding redshift is given by the expression 
\begin{small}
\begin{equation}
Z_{s}=\frac{1}{\sqrt{1-2u}}-1=\frac{1}{\sqrt{1+r\Lambda -\frac{3ar\left[
1+C\left( 1+4ar^{2}\right) ^{-\frac{1}{2}}\right] }{2\left( 1+ar^{2}\right) }
}}-1  \label{eq:30}
\end{equation}
\end{small}
According to Buchdahl \cite{Buchdahl1959}, the maximum value of $u(r)$ i.e. $
\left( \frac{m(r)}{r}\right) _{max}$ is $\frac{4}{9}$ in $(3+1)$ dimension. The maximum allowed value of surface redshift \cite{Haensel2000} is $Z_{s}\leq 0.85$ in $(3+1)$ dimension. Cruz and Zanelli \cite{Cruz:1994ar} have
shown that Buchdahl’s theorem \cite{Buchdahl1959} remains to uphold also in $(2 + 1)$ dimensions. Therefore, we will use the same upper limit of the ratio of the mass to radius i.e. $\frac{m(r)}{r}\leq \frac{4}{9}$ and the surface redshift $Z_{s}\leq 0.85$ in $(2+1)$ dimensions as we used in the case of $(3+1)$ dimension. 
\subsection{Generalized TOV Equation}

The generalized TOV equation for anisotropic system is written as 
\begin{equation}
\frac{d}{dr}\left( p_{r}-\frac{\Lambda }{2\pi }\right) +\nu
^{\prime }(\rho +p_{r})+\frac{1}{r}(p_{r}-p_{t})=0.  \label{eq:31}
\end{equation}
This equation represents the equilibrium condition of the system under
gravitational force ($F_{g}$), hydrostatic force ($F_{h}$) and anisotropic force 
($F_{a}$) as

\begin{equation}
F_{h}+F_{g}+F_{a}=0,  \label{eq:32}
\end{equation}
where 
\begin{align}
F_{g}=-\nu ^{\prime }(\rho +p_{r}), && F_{a}=\frac{1}{r}(p_{t}-p_{r})  \nonumber\\ 
 \text{and} && F_{h}=-\frac{d}{dr}\left( p_{r}-\frac{\Lambda }{2\pi }\right)\nonumber
\end{align}
The following form of the gravitational force ($F_{g}$), hydrostatic force ($F_{h}$) and anisotropic force ($F_{a}$) for the Heintzmann line element is given below.
\begin{small}
\begin{eqnarray}
&&F_{g}=\frac{9a^{2}r}{4\pi \left(
ar^{2}+1\right) ^{3}\left( 4ar^{2}+1\right) ^{3/2}}  \nonumber \\
&&\times \left[ C\left( 14a^{2}r^{4}+ar^{2}-1\right) \right.   \nonumber \\
&&\left. +\left( 4ar^{2}+1\right) ^{\frac{1}{2}}\left(
4a^{2}r^{4}-11ar^{2}-3\right) \right] ,  \label{eq:33}
\end{eqnarray}
\begin{eqnarray}
&&F_{h}=\frac{3a^{2}r}{2\pi \left( ar^{2}+1\right) ^{3}\left( 4ar^{2}+1\right)
^{3/2}}  \nonumber \\
&&\times \left[ C\left( -18a^{2}r^{4}+3ar^{2}+3\right) \right.   \nonumber \\
&&\left. +\left( 4ar^{2}+1\right) ^{\frac{1}{2}}\left(
-4a^{2}r^{4}+19ar^{2}+5\right) \right] ,  \label{eq:34}
\end{eqnarray}%
\begin{equation}
F_{a}=-\frac{3a^{2}r\left[ C\left( 6ar^{2}+3\right) +\left( 4ar^{2}+1\right)
^{3/2}\right] }{4\pi \left( ar^{2}+1\right) ^{2}\left( 4ar^{2}+1\right)
^{3/2}}.  \label{eq:35}
\end{equation}
\end{small}

\section{Conditions on the Metric parameters}

Eqs. (\ref{eq:25}-\ref{eq:27}) indicate that both energy density and
pressure are monotonically decreasing function and get maximized at the
centre for any values of $a$ and $C$ in a positive domain. The central
pressure and $A$ will be positive and real only when satisfying the
conditions 
\begin{equation}
-3<3M_{0}+2\Lambda R^{2}<0.  \label{eq:36}
\end{equation}
The positivity of $C$ implies that 
\begin{equation}
M_{0}\left( 4\Lambda R^{2}+2\right) +2M_{0}^{2}+\Lambda R^{2}\left( 2\Lambda
R^{2}+1\right) <0.  \label{eq:37}
\end{equation}
The dominating energy condition at the centre yields 
\begin{eqnarray}
&&3\sqrt{3}\sqrt{\frac{M_{0}+2\Lambda R^{2}}{3M_{0}+2\Lambda R^{2}}} 
\nonumber  \label{eq:38} \\
&&\times \left[ M_{0}\left( 4\Lambda R^{2}+2\right) +2M_{0}^{2}+\Lambda
R^{2}\left( 2\Lambda R^{2}+1\right) \right]   \nonumber \\
&&-3\Lambda R^{2}-4\Lambda R^{2}\left( 3M_{0}+2\Lambda R^{2}\right) \Lambda
>0.
\end{eqnarray}
The strong energy condition at the surface implies that 
\begin{equation}
-\frac{\Lambda M_{0}R^{2}-M_{0}^{2}-M_{0}+4\Lambda ^{2}R^{4}+\Lambda R^{2}}{
2\pi M_{0}R^{2}+4\pi \Lambda R^{4}}\geq 0.  \label{eq:39}
\end{equation}

Table-\ref{tab:app2} and \ref{tab:app1} present the analytic and numerical
form of solutions of these above constrained equations \ref{eq:36}-\ref{eq:39} along with Buchdahl condition \cite{Buchdahl1959} $M<\frac{4}{9}R$,
respectively.

\section{Example}

\subsection{PSR B0943+10}

Yue et al., in the paper \cite{Yue2006} showed that the small polar gap of PSR B0943+10 could be explained using Ruderman-Sutherland-type vacuum gap model if the pulsar has the mass and radii about $0.02M_{\odot }$ and $2.6$ km respectively. We use our model for this pulsar and find out the useful parameters. If we use the mass of this pulsar as a only input parameter, then for $\Lambda =-0.001$, $\Lambda =-0.005$ and $\Lambda =-0.01$ our model predicts that the radii of this pulsar is in the range $0.066447<R<6.25384$,  $0.066447<R<2.7968$ and $0.066447<R<1.97764$ respectively. We enlist some parameters of this pulsar calculated using our model in table-\ref{tab:2}. One can observe (Fig.-\ref{fig:2} and \ref{fig:3}) that matter-energy density, radial pressure and transverse pressure are maximum at the centre and decreases monotonically towards the boundary. Also, one can see that radial pressure drops to zero at the boundary, while density does not. And although the energy density monotonically decreases, its value remains very high throughout the stellar system. Therefore, it may be justified to take these compact stars as strange stars where the surface density remains finite rather than the neutron stars for which the surface density vanishes at the boundary \cite{Haensel1986,Alcock1986,Farhi1984,Dey1998,PhysRevD.84.083002,Norman}. Fig.\ref{fig:4} suggests that the anisotropic force is attractive for our model. The attractiveness of the anisotropic force disallows the formation of massive compact star \cite{Kalam2012}. In other words, it allows the formation of the low mass compact star (strange star for our case). Fig.\ref{fig:8} ensures that all the energy conditions are satisfied in our model. It is apparent from Fig.\ref{fig:5} that the compactness, $u(r)=\frac{m(r)}{r}$ is an increasing function of the radial parameter and has maximum value below $\frac{4}{9}$ for all six cases. According to Buchdahl \cite{Buchdahl1959}, the maximum value of $u(r)$ i.e. $\left( \frac{m(r)}{r}\right) _{max}$ is $\frac{4}{9}$ in $(3+1)$ dimension. Fig \ref{fig:7} shows that the value of surface gravitational red-shift has much less value than the maximum allowed value ($Z_{s}\leq 0.85$) \cite{Haensel2000} in $(3+1)$ dimension. Fig.\ref{fig:9} shows that the gravitational force$(F_{g})$, the hydrostatic force$(F_{h})$ and the anisotropic force$(F_{a})$ are in equilibrium in the interior region of the star. 
\begin{table*}[!htbp]
\small
\begin{tabular}{@{}p{1.8cm}p{1.1cm}p{1.2cm}p{1.2cm}p{1.4cm}p{1.7cm}p{1.5cm}p{1.5cm}p{1.6cm}p{1.5cm}}
\hline
Cosmological constant($\Lambda$) & Radius (R) in km & Central energy density
($\rho _0$) in $10^{14}\frac{g}{cc}$ & Surface energy density ($\rho _R$) in $%
10^{14}\frac{g}{cc}$ & Central pressure ($p_0$) in $10^{33}\frac{dyne}{cm^{2}}$ & 
Compactness $(u)$ & Surface Redshift $Z_{s}$ & A & a & c \\ \hline
& 1 & 126.757 & 126.495 & 11.9498 & 0.029532 & 0.0309081 & 0.970019 & 
0.00035401 & 108.462 \\ 
-0.001 & 2.6 & 18.8449 & 18.6192 & 11.1187 & 0.011358 & 0.0115557 & 0.970027
& 0.00035257 & 13.7302 \\ 
& 6.25 & 3.25651 & 3.22694 & 6.58000 & 0.004725 & 0.0047589 & 0.970285 & 
0.00034472 & 0.00369 \\[1ex] 
& 0.7 & 259.054 & 257.788 & 58.6996 & 0.042189 & 0.0450611 & 0.970020 & 
0.00176822 & 42.6728 \\ 
-0.005 & 1.5 & 56.7741 & 55.7879 & 52.3773 & 0.019688 & 0.0202892 & 0.970041
& 0.00175728 & 7.14971 \\ 
& 2.79 & 16.3438 & 16.1918 & 32.9976 & 0.010585 & 0.0107560 & 0.970283 & 
0.00172375 & 0.01464 \\[1ex] 
& 0.5 & 507.770 & 505.241 & 117.327 & 0.059064 & 0.0648715 & 0.970020 & 
0.00353631 & 41.7653 \\ 
-0.01 & 1 & 127.657 & 125.606 & 106.541 & 0.029532 & 0.0309081 & 0.970037 & 
0.00351765 & 8.38975 \\ 
& 1.97 & 32.7843 & 32.4739 & 66.1483 & 0.014991 & 0.0153366 & 0.970281 & 
0.00344777 & 0.02329 \\ \hline
\end{tabular}%
\caption{Value of some parameters of the Pulsar PSR B0943+10 estimated using
our model}
\label{tab:2}
\end{table*}

\subsection{PSR J1640-4631}

The X-ray pulsar PSR J1640-4631 high breaking index was explained by Chen 
\cite{Chen2016} by considering the pulsar as a low mass neutron star having
mass about $0.1M_{\odot}$. Chen \cite{Chen2016} showed that the radius of
this pulsar is about 29 km using the formula $R \propto M^{-1/3} $ . Using
our model, we get that the possible radius ranges for this pulsar are $%
0.332235 < R < 13.7623$, $0.332235 < R < 6.15469$ and $0.332235 < R < 4.35202
$ for $\Lambda=-0.001$, $\Lambda=-0.005$ and $\Lambda=-0.01$ respectively.
In this paper, we estimate some parameters of this pulsar using our model
presented in table-\ref{tab:3}. All the graphs related to this pulsar is
very similar to the Pulsar PSR B0943+10 (we do not present graphs of this
pulsar in this paper).

\begin{table*}[!htbp]
\small
\begin{tabular}{@{}p{1.8cm}p{1.1cm}p{1.2cm}p{1.2cm}p{1.4cm}p{1.7cm}p{1.5cm}p{1.5cm}p{1.6cm}p{1.5cm}}
\hline
Cosmological constant($\Lambda$) & Radius (R) in km & Central energy density
($\rho _0$) in $10^{14}\frac{g}{cc}$ & Surface energy density ($\rho _R$) in $%
10^{14}\frac{g}{cc}$ & Central pressure ($p_0$) in $10^{33}\frac{dyne}{cm^{2}}$ & 
Compactness $(u)$ & Surface Redshift $Z_{s}$ & A & a & c \\ \hline
& 1 & 634.023 & 632.238 & 80.4903 & 0.147660 & 0.1912530 & 0.839452 & 
0.00047258 & 414.783 \\ 
-0.001 & 7 & 13.5744 & 12.3067 & 68.6351 & 0.021094 & 0.0217861 & 0.840080 & 
0.00045207 & 6.86262 \\ 
& 13.76 & 3.43523 & 3.25908 & 39.2116 & 0.010731 & 0.0109070 & 0.847190 & 
0.00040117 & 0.00098 \\[1ex] 
& 1 & 637.489 & 628.800 & 397.307 & 0.147660 & 0.1912530 & 0.839459 & 
0.00235400 & 81.7961 \\ 
-0.005 & 3 & 73.7089 & 67.1735 & 347.918 & 0.049220 & 0.0531802 & 0.839985 & 
0.00226856 & 7.63430 \\ 
& 6.15 & 17.1999 & 16.3122 & 196.267 & 0.024010 & 0.0249106 & 0.847174 & 
0.00200621 & 0.00448 \\[1ex] 
& 0.5 & 2541.37 & 2523.70 & 801.036 & 0.295320 & 0.5629580 & 0.839454 & 
0.00471912 & 165.049 \\ 
-0.01 & 2 & 165.243 & 151.671 & 707.789 & 0.073830 & 0.0831624 & 0.839876 & 
0.00455780 & 8.81130 \\ 
& 4.35 & 34.3760 & 32.6076 & 392.325 & 0.033945 & 0.0357772 & 0.847182 & 
0.00401205 & 0.00273 \\ \hline
\end{tabular}%
\caption{Value of some parameters of the Pulsar PSR J1640-4631 estimated
using our model}
\label{tab:3}
\end{table*}

\subsection{1E 1207.4-5209}

Xu in a paper \cite{Xu2005} showed that the radio-quiet object 1E
1207.4-5209 could be a low mass bare strange star with mass and radii $%
10^{-3}M_{\odot }$ and $1$ km respectively. The radius ranges predicted by
our model are $0.00332235<R<1.40291$, $0.00332235<R<0.6274$ and $%
0.00332235<R<0.443639$ for $\Lambda =-0.001$, $\Lambda =-0.005$ and $\Lambda
=-0.01$, respectively. Table-\ref{tab:4} presents some parameters of this
pulsar calculated using our model. Graphs related to this pulsar are similar
to the pulsar PSR J1640-4631 (we do not show graphs of this pulsar in this
paper).

\begin{table*}[!htbp]
\small
\begin{tabular}{@{}p{1.8cm}p{1.1cm}p{1.2cm}p{1.2cm}p{1.4cm}p{1.7cm}p{1.5cm}p{1.5cm}p{1.6cm}p{1.5cm}}
\hline
Cosmological constant($\Lambda$) & Radius (R) in km & Central energy density
($\rho _0$) in $10^{14}\frac{g}{cc}$ & Surface energy density ($\rho _R$) in $%
10^{14}\frac{g}{cc}$ & Central pressure ($p_0$) in $10^{33}\frac{dyne}{cm^{2}}$ & 
Compactness $(u)$ & Surface Redshift $Z_{s}$ & A & a & c \\ \hline
& 0.5 & 25.3205 & 25.3093 & 0.53816 & 0.002952 & 0.0029651 & 0.998523 & 
0.00033426 & 20.5610 \\ 
-0.001 & 1 & 6.33220 & 6.32525 & 0.44135 & 0.001476 & 0.0014793 & 0.998523 & 
0.00033410 & 2.89830 \\ 
& 1.4 & 3.22967 & 3.22821 & 0.31754 & 0.001054 & 0.0010560 & 0.998524 & 
0.00033388 & 0.01126 \\[1ex] 
& 0.1 & 632.903 & 632.841 & 2.82006 & 0.014760 & 0.0150950 & 0.998523 & 
0.00167155 & 114.747 \\ 
-0.005 & 0.3 & 70.3539 & 70.3034 & 2.56211 & 0.004921 & 0.0049573 & 0.998523
& 0.00167110 & 10.0970 \\ 
& 0.627 & 16.1019 & 16.0948 & 1.58408 & 0.002354 & 0.0023624 & 0.998524 & 
0.00166941 & 0.00261 \\[1ex] 
& 0.1 & 632.933 & 632.812 & 5.57552 & 0.014760 & 0.0150950 & 0.998523 & 
0.00334298 & 55.8811 \\ 
-0.01 & 0.2 & 158.292 & 158.188 & 5.18879 & 0.007381 & 0.0074637 & 0.998523
& 0.00334231 & 11.7326 \\ 
& 0.44 & 32.6976 & 32.6817 & 3.20637 & 0.003354 & 0.0033715 & 0.998524 & 
0.00333888 & 0.04858 \\ \hline
\end{tabular}%
\caption{Value of some parameters of the Pulsar 1E 1207.45209 estimated
using our model}
\label{tab:4}
\end{table*}

\section{Discussion and Conclusions}

In this paper, we have presented a new model of the anisotropic low mass
strange stars based on the Heintzmann ansatz in $(2+1)$ dimension. The
attractive anisotropic force plays a significant role to bound the upper
mass limit(which is comparatively low) of the strange star. The upper mass
limit of the strange star for our model comes out as $M(R)_{max}<0.536438$ $%
km$ or $0.3633M_{\odot }$ for $\Lambda >-0.332447km^{-2}$ and $%
M(R)_{max}<M_{1}$ $km$ for $\Lambda \leq -0.332447km^{-2}$, where $M_{1}$
(see Eq-\ref{M1}) depends on cosmological constant $\Lambda $. So, we can
say that "Strange stars, if they exist, can play an important role in the
solution to the cosmological constant problem "\cite{Bambi2007,Kalam2013}.

Table-\ref{tab:app2} presents the possible range of the cosmological
constant $\Lambda $, radius ($R$) and mass ($M$) of the low mass strange
star. Fig-\ref{fig:1} shows the accessible mass-radius region of our model.
In this available region of the mass and radius, Buchdahl condition, all the
energy conditions and other boundary conditions of the compact star are
satisfied and the metric parameters $A$, $a$ and $C$ are real and positive.

This model can be useful to analyze the low mass bare strange star \cite%
{Xu2005}. In the paper \cite{Yue2006}, authors showed that PSR B0943+10 is a
strange star having mass $\sim 0.02M_{\odot }$ and radius $\sim 2.6$ km. Our
model predict that a pulsar of mass $0.02M_{\odot }$ has radii in between $%
0.066447<R<6.25384$, $0.066447<R<2.7968$ and $0.066447<R<1.97764$ for $%
\Lambda =-0.001$, $\Lambda =-0.005$ and $\Lambda =-0.01$ respectively. Chen 
\cite{Chen2016} argued that the high braking index of the PSR J1640-4631 can
be explained by considering that PSR J1640-4631 is a low mass neutron star
of mass $0.1M_{\odot }$. Using our model, we obtain that PSR J1640-4631 is a
strange star because of non-vanishing surface energy density and get the
radius range $0.332235<R<13.7623$, $0.332235<R<6.15469$ and $%
0.332235<R<4.35202$ for $\Lambda =-0.001$, $\Lambda =-0.005$ and $\Lambda
=-0.01$, respectively. Xu \cite{Xu2005} showed that 1E 1207.4\^{a}
 could have radius $1$ km and mass around $10^{-3}M_{\odot }$. From our
model, we get radius range $0.00332235<R<1.40291$, $0.00332235<R<0.6274$ and 
$0.00332235<R<0.443639$ for $\Lambda =-0.001$, $\Lambda =-0.005$ and $%
\Lambda =-0.01$, respectively. 

As the cosmological constant decreases, the central energy density and
pressure and surface energy density increase for all three cases. For the
fixed cosmological constant, the central energy density and pressure and
surface energy density decrease as the radius increase in all three cases
(see table:\ref{tab:2}-\ref{tab:4}). Fig.-\ref{fig:9} indicates that the
gravitational force $F_{g}$ and anisotropic force $F_{a}$ become identical
while compactness is approaching the Buchdahl limit $\frac{4}{9}$ \cite%
{Buchdahl1959}.

In this paper, we find the EoS having form $p_{r}=\alpha \rho +\beta$, where $\alpha$ is a dimensionless constant and $\beta$ has a dimension of $km^{-2}$ is also a constant. Fig.-\ref{fig:10} shows the relation between  stellar interior radial pressure and energy density. Fig.-\ref{fig:10} indicates that EoS found in our model is on the softer side.

It is to be mentioned here that we are comparing our results with the data
from a $3+1$ dimensional object. Though for an observer in the $\theta =\pi
/2$ or $const.$ plane, the measured data will be approximately the same for
both dimensions.

\section*{Acknowledgments}

MM is thankful to CSIR (Grand No.-09/1157(0007)/ 2019-EMR-I) for providing
financial support. NR is thankful to UGC MANF(MANF-2018-19-WES-96213) for
providing financial support. MK is grateful to the Inter-University Centre
for Astronomy and Astrophysics (IUCAA), Pune, India for providing
Associateship programme under which a part of this work was carried out. 
\section*{Appendix:A}
\label{app}
\begin{small}

\begin{eqnarray}  \label{M1}
&&M_{1}=-\frac{64}{243 \Lambda }+  \nonumber \\
&&\frac{4 (729 \Lambda +128)}{243 \left(54 \sqrt{-98304\Lambda ^{7}-4374
\Lambda ^{8} (243 \Lambda +128)}+4096 \Lambda ^3\right)^{\frac{1}{3}}} 
\nonumber \\
&&+\frac{2 \left(54 \sqrt{-98304\Lambda ^{7}-4374 \Lambda ^{8} (243 \Lambda
+128)}+4096 \Lambda ^3\right)^{\frac{1}{3}}}{243 \Lambda ^2}  \nonumber \\
\end{eqnarray}

\begin{equation}
R_{1}=\frac{1}{2} \sqrt{\frac{3}{\Lambda }+\frac{\sqrt{\Lambda ^2 (9-16 M)}}{%
\Lambda ^2}-\frac{8 M}{\Lambda}}
\end{equation}

\begin{equation}
R_{3}=\frac{1}{2} \sqrt{-\frac{M}{\Lambda }-\frac{\sqrt{M (17 M-8)}}{\Lambda 
}}
\end{equation}

\begin{equation}
R_{4}=\frac{9}{8} \sqrt{M+\sqrt{M (17 M-8)}}
\end{equation}

\begin{equation}
R_{5}=\frac{9}{8} \sqrt{8 M+\sqrt{9-16 M}-3}
\end{equation}

\begin{equation}
R_{7}=\frac{1}{2} \sqrt{\frac{\sqrt{M (17 M-8)}}{\Lambda }-\frac{M}{\Lambda }%
}
\end{equation}

\begin{eqnarray}
M_{3}(x)=128 - 256 x + 648 \Lambda x^2 + 6561 \Lambda^2 x^3
\end{eqnarray}

\begin{eqnarray}  \label{s(x)}
&&R_{2}(x)=22 \Lambda ^5 x^{10}+ \left(198 \Lambda ^4 M-69 \Lambda
^4\right)x^8  \nonumber \\
&&+ \left(54 \Lambda ^3+864 \Lambda ^3 M^2-522 \Lambda ^3 M\right)x^6 
\nonumber \\
&&+ \left(2160 \Lambda ^2 M^3-1944 \Lambda ^2 M^2+432 \Lambda ^2 M\right)x^4
\nonumber \\
&&+ \left(2592 \Lambda M^4-3240 \Lambda M^3+1296 \Lambda M^2-162 \Lambda
M\right)x^2  \nonumber \\
&&+864 M^5-1296 M^4+648 M^3-108 M^2
\end{eqnarray}
\begin{eqnarray}  \label{t(x)}
&&M_{2}(x)=1420541793 \Lambda ^5 x^8+2525407632 \Lambda ^4 x^7  \nonumber \\
&&+\left(2176782336 \Lambda ^3-880066296 \Lambda ^4\right) x^6  \nonumber \\
&&+\left(1074954240 \Lambda ^2-1315139328 \Lambda ^3\right) x^5  \nonumber \\
&&+\left(136048896 \Lambda ^3-967458816 \Lambda ^2+254803968 \Lambda \right)
x^4  \nonumber \\
&&+\left(214990848 \Lambda ^2-318504960 \Lambda +16777216\right) x^3 
\nonumber \\
&& +(127401984 \Lambda -25165824) x^2  \nonumber \\
&&+(12582912-15925248 \Lambda ) x-2097152
\end{eqnarray}
\end{small}
\begin{small}
\begin{eqnarray}
&&R_{6}(x)=94143178827 M^2-564859072962 M^3  \nonumber \\
&& +1129718145924 M^4-753145430616 M^5  \nonumber \\
&&+\left(-27894275208 M+223154201664 M^2\right.  \nonumber \\
&& \left.-557885504160 M^3+446308403328 M^4\right) x^2  \nonumber \\
&&-x^4 \left(14693280768 M-66119763456 M^2+73466403840 M^3\right)  \nonumber
\\
&&+\left(362797056-3507038208 M+5804752896 M^2\right) x^6  \nonumber \\
&&+(91570176-262766592 M) x^8+5767168 x^{10}
\end{eqnarray}
\end{small}

\bibliographystyle{hindawi_bib_style}  
\bibliography{2+1_AHEP}                

\end{document}